\documentclass[12pt]{iopart}
\usepackage{graphicx}
\usepackage{psfrag}
\begin{document}
\title[Band gap effects in graphene due to hydrogen adsorption]
{First-principles study of bandgap effects in graphene due to hydrogen
adsorption}
\author{Mahboobeh Mirzadeh and Mani Farjam}
\address{School of Nano-Science, Institute for Research in Fundamental
Sciences (IPM), PO~Box 19395-5531, Tehran, Iran}
\begin{abstract}
Hydrogen adsorption on graphene in commensurate periodic arrangements
leads to
bandgap opening at the Dirac point and the emergence of dispersionless
midgap bands.
We study these bandgap effects and their dependence on periodicity
for a single hydrogen adsorbate
on periodic graphene supercells
using spin-polarized density-functional theory calculations.
Our results show that for certain periodicities,
marked by a scale factor of three, 
the bandgap is suppressed to a great extent,
and has a special level structure around the neutrality point.
We present explanations for the origin of the
changes to the band structure in terms of
the \textit{ab initio} Hamiltonian matrix.
This method may be used to obtain a more accurate tight-binding description
of single hydrogen adsorption on graphene.
\end{abstract}
\submitto{\JPCM}
\section{Introduction}

The isolation/discovery of graphene
\cite{novo2004},
a single layer of graphite,
and the demonstration of its unusual electronic properties
\cite{novo2005, zhan2005}
has attracted a lot of attention in the condensed matter physics community.
The origin of the novel phenomena which were observed in these works
lies in the special band structure
of graphene that was studied theoretically several decades ago
\cite{wall1947}.
The theoretical analysis
showed that the valence and conduction bands in graphene
touch at only the two inequivalent corners of
the hexagonal Brillouin zone (BZ),
known as the Dirac points.
The energy dispersion in the neighbourhood of
these points is linear,
and the energy surfaces are in the shape of Dirac cones.
The implication of all this is that the charge carriers would
behave as two-dimensional (2D) massless chiral Dirac fermions
\cite{cast2009},
and possess their rich physical properties.
Recently device engineers are also showing interest in graphene,
because of its exceptional two-dimensional character and high-carrier mobility
\cite{schw2010}.
However, the zero-gap spectrum renders graphene a semimetal, which is
a limitation for applications in logic circuits.

Hydrogen interaction with graphene is of diverse interest,
and has been the subject of many theoretical and experimental
investigations.
As recent studies have shown, hydrogen functionalization may induce
a metal-insulator transition in graphene
\cite{elia2009}
and open a bandgap in its electronic spectrum
\cite{balo2010, habe2010, habe2011a}.
Another interesting property, with potential applications in spintronics,
is the emergence of magnetism in graphene materials
due to interaction with hydrogen
\cite{yazy2010}.
In addition,
hydrogen adsorbates are believed to be a dominant source of scattering
in graphene that limits its electron mobility
\cite{ni2010,kato2010}.
These and other fascinating properties provide the motivation
to study and understand hydrogen interaction with graphene.

\begin{figure}
\psfrag{(a)}{\bf (a)}
\psfrag{(b)}{\bf (b)}
\centering
\includegraphics[width=0.4\linewidth]{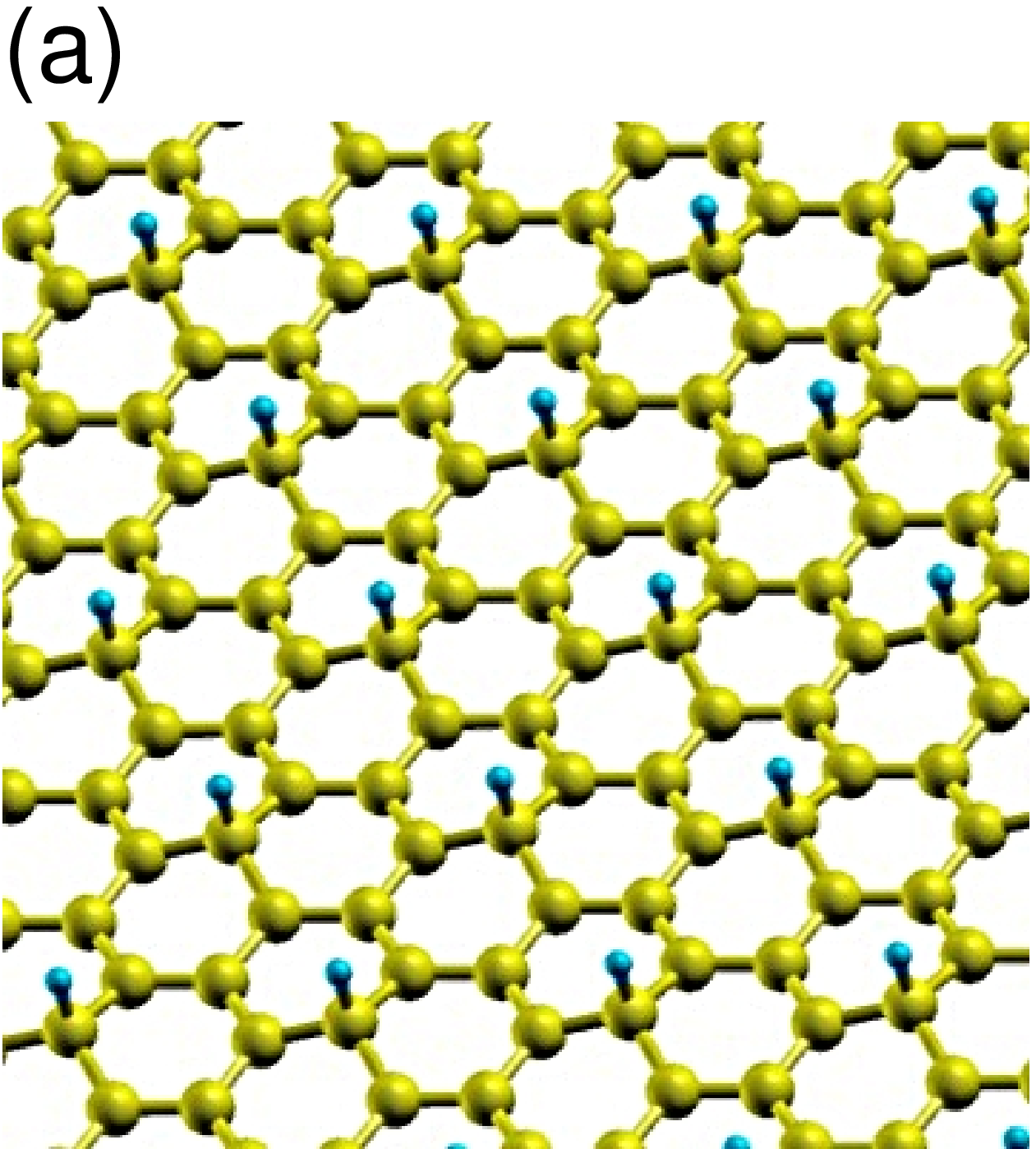} \\
\vspace{5mm}
\includegraphics[width=0.4\linewidth]{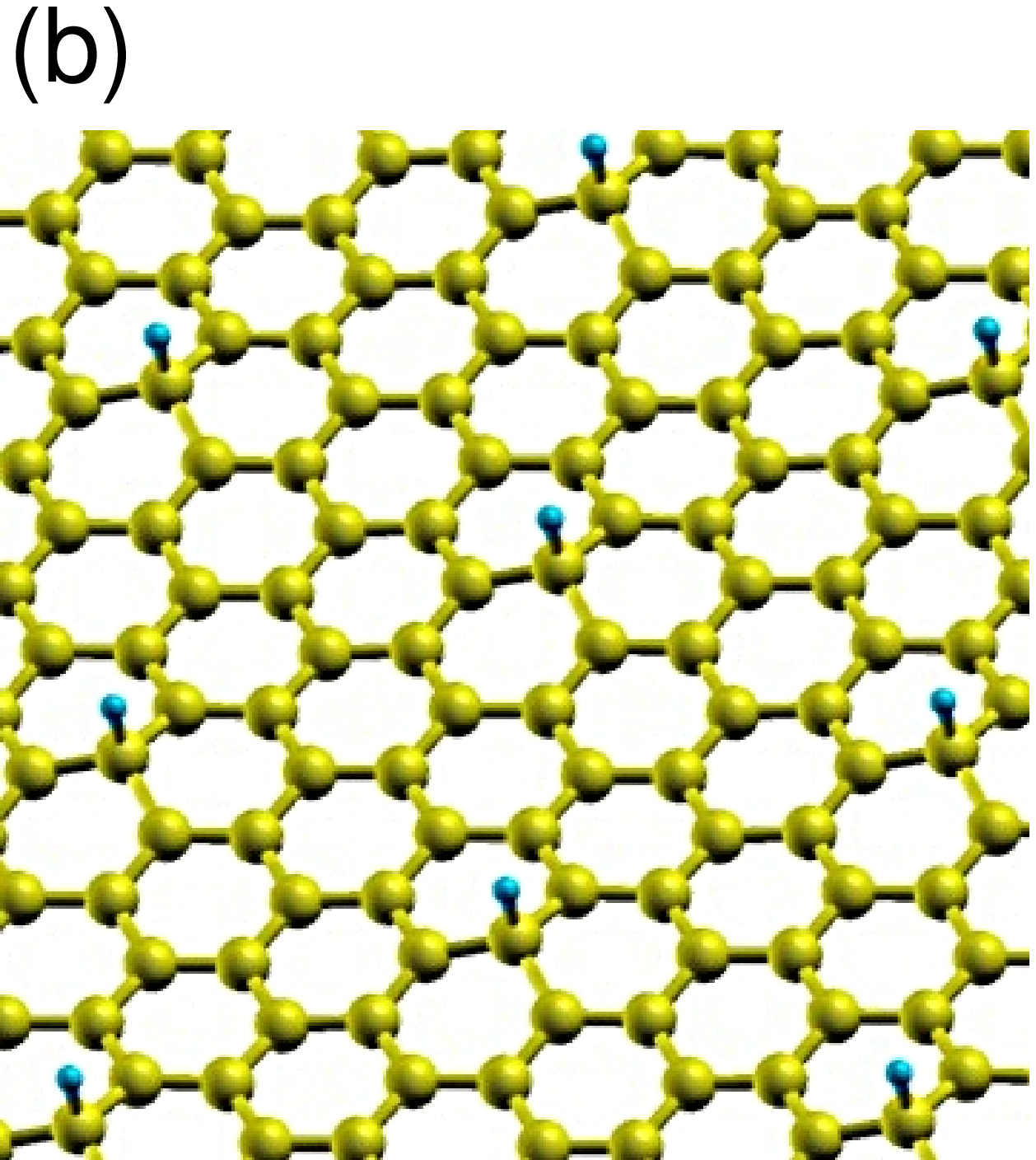}
\caption{\label{fig1}
Hydrogen chemisorbs on top of a C atom causing it to move out of plane
to accomodate the rehybridized bond.
(a) $\sqrt{3}\times\sqrt{3}R30^\circ$ coverage corresponding to
HC$_6$, and (b) $3\times3$ coverage corresponding to HC$_{18}$.}
\end{figure}

Much of the understanding of hydrogen adsorption on graphene has come
through density-functional theory (DFT) calculations
\cite{jelo1999, dupl2004, yazy2007, bouk2008, caso2009, verg2010}.
Thus it has been shown that hydrogen chemisorbs on graphene
in an atop position, causing partial sp$^2$ to sp$^3$
rehybridization, and creating a p$_z$ orbital vacancy in the process
(figure~\ref{fig1}).
This implies that hydrogen and vacancy defects have some common properties,
such as magnetic moments and resonant scattering.
The effect on the electronic structure, calculated by DFT on $n\times{n}$
graphene supercells with a single hydrogen adsorbate,
is the appearance of a pair
of spin-polarized states within a bandgap opened around the Dirac point.
Other changes include the appearance of
spin densities in the graphene plane in a characteristic
$\sqrt{3}\times\sqrt{3}R30^\circ$ texture centred on the adsorption site
which can be resolved
by scanning tunneling microscopy (STM)
\cite{yazy2007}.
The midgap state is a universal feature of resonant scatterers
shared by other covalently-bonded adsorbates, including CH$_3$, NH$_2$,
OH, and F
\cite{wehl2010},
making H/graphene a prototype of a wider class of systems.
DFT calculations have been supplemented by semiempirical
tight-binding (TB) model to
extend to larger number of atoms
than can be handled in first-principles calculations.
This model has been used in the study of an isolated hydrogen adsorbate
\cite{farj2011}
and random distributions of hydrogen adsorbates
\cite{yuan2010}.

\begin{figure}
\centering
\includegraphics[width=0.5\linewidth]{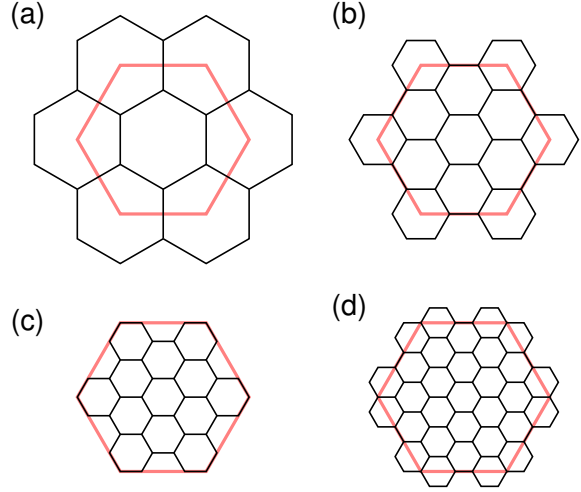}
\caption{\label{fig2}
Relationship between Brillouin zones of graphene (red) and those of its various
superstructures (black).
Brillouin zones of (a) $\sqrt{3}\times\sqrt{3}R30^\circ$, (b) $3\times3$,
(c) $4\times4$, and (d) $5\times5$ superstructures.
In (a) and (b) both K points (corners) of the
BZ of graphene coincide with the $\Gamma$ point (centre)
of the BZ of the superstructure.
In (c) and (d) the K points of graphene coincide with the K points of the
BZ of the superstructure.}
\end{figure}

Recently, it was shown that bandgap opening has a strong dependence
on the periodicity of the hydrogen arrangement on graphene
\cite{garc2010},
so that the bandgap is much smaller for certain periodicities than expected.
The existence of different cases with respect to bandgap opening
can be explained on the basis of
the geometrical relationship between the Brillouin zones of graphene
and its periodic superstructures
(figure \ref{fig2}),
although the precise changes to the band structures would require
numerical calculations.
In this paper, we focus on DFT calculations of the band structure of graphene
with a single hydrogen adsorbate on periodically repeated unit cells
in order to study the predicted dependence of bandgap effects
on periodicity.
We find that for the special periodicities the bandgap 
is suppressed to a great extent,
particularly when spin polarization is neglected.
Within spin-polarized calculation, the bandgap is somewhat wider,
and has a distinct and notable structure,
not predictable within the commonly used simple tight-binding model.
We attempt to explain the origin of
these effects based on
tight-binding models extracted from {\it ab initio} Hamiltonian matrices.
This approach is similar to the one used for an accurate
tight-binding description of graphene
\cite{reic2002}.
We remark that the alternative semiempirical TB parameters,
obtained by fitting to DFT derived band structures, 
turn out to be very different from the {\it ab initio} parameters.
It seems, therefore, necessary to compare more closely
the results obtained from semiempirical models and
first-principles calculations wherever possible.

The rest of the paper is organized as follows.
In section 2 we describe the computational details. In section 3 we present
the results and discuss their salient features. In the final section we
present a summary and our conclusions.

\section{Computational method}
For the choices of $n\times{n}$ and
$n\sqrt{3}\times{n\sqrt{3}}R30^\circ$ unit cells
of graphene with a hydrogen atom adsorbed on top of one of the carbon atoms,
as shown in figure 1, we obtain the electronic structure using DFT calculations.
For all the calculations we used the \textsc{Siesta} DFT code,
an efficient code that allows us
to study relatively low adsorbate coverages of H/graphene,
with supercells of up to a few hundred carbon atoms.
The calculations use a basis set of
numerical atomic orbitals
\cite{sole2002},
norm-conserving pseudopotentials
\cite{trou1991},
and the generalized gradient approximation (GGA) of exchange-correlation
density functional
\cite{perd1996}.
The use of localized atomic orbitals in \textsc{Siesta} makes it
convenient to extract tight-binding matrix elements from
the Hamiltonian matrix.
Our calculations use
the diagonalization-based method of solving the Kohn-Sham equations and
include both restricted and unrestriced spins.
We chose a separation of $20$ \AA\ between repeated slabs,
to represent the 2D systems,
and for $k$-point sampling we used various 
Monkhorst-Pack grids
\cite{monk1976}
typically corresponding to $36\times36\times1$ points in the
graphene Brillouin zone ($\Gamma$ included).
The systems were fully relaxed,
until the atomic forces were smaller than 0.01 eV/\AA, before
calculation of the band structure and density of states.
We used basis sets of double-$\zeta$ plus polarization (DZP) quality for most
of the results reported, and also single-$\zeta$ (SZ) quality,
to reduce computing time, for some of the
data involving larger systems.

\section{Results}
Our systems consist of periodically repeated supercells containing
one hydrogen adsorbate and an integral
number of graphene unit cells
(figure 1).
This configuration is commonly used to describe single hydrogen
adsorption on an infinite graphene, under the assumption that if the
$n\times{n}$ supercell is large enough, say $n\ge4$, H adsorbates
may be considered as isolated from each other.
Various properties, however, do depend on the scale
of the supercell, as we demonstrate in this section.

Figure 1 shows two samples of hydrogenated graphenes,
$\sqrt{3}\times\sqrt{3}R30^\circ$ and $3\times3$,
which show the periodic arrangement and also happen to
represent special periodicities, to be explained below.
Since it has been established that covalently bonded hydrogen chemisorbs on
top of a carbon atom we simply used this placement in our calculations,
and allowed our DFT code to perform structural relaxation to find the bond
lengths and angles.
We found results in agreement with previous calculations,
where the H-C bond is 1.13 \AA, and the C-C bond elongates from 1.42 \AA\ to
1.51 \AA\ as a result of sp$^2$ to sp$^3$ rehybridization.
The bond angles do not form the perfect tetrahedral angles of 109.5$^\circ$,
but have a mixed character between the two types of hybridization and,
furthermore, they may depend on hydrogen concentration.

Our focus is on the changes made to the band structure
as a result of hydrogen adsorption,
and on the understanding of their underlying causes.
Toward this end it is useful to compare with
the simple spinless tight-binding model
that describes hydrogen adsorption on graphene
and includes only $\pi$ bands,
\begin{equation}
\mathcal{H}=-\gamma_\mathrm{CC}\sum_{\langle{i,j}\rangle} c_i^\dag c_j
+\varepsilon_\mathrm{H} d^\dag d
+\gamma_\mathrm{HC} (c^\dag_0 d + d^\dag c_0).
\end{equation}
Typical values are $\gamma_\mathrm{CC}=2.7$~eV for hopping between the carbon
2p$_z$ orbitals,
$\varepsilon_\mathrm{H}=-0.2$~eV for the hydrogen
1s orbital on-site energy relative to that of carbon 2p$_z$,
and $\gamma_\mathrm{HC}=-5.2$~eV for hopping between H and C orbitals
\cite{wehl2010}.
This is a semiempirical model whose
parameters were derived by fitting particular DFT band structure
calculations, and it provides a useful description of
the midgap resonance.

If we use the tight-binding model given by (1) to calculate the
band structures for $4\times4$ or $5\times5$ hydrogenated
graphenes we find bandgap openings at their respective K points and an almost
dispersionless midgap band.
On the other hand, if we use it to calculate the band structure for 
$\sqrt{3}\times\sqrt{3}R30^\circ$ and $3\times3$ coverages no bandgap opens,
while the degeneracy point shifts to the $\Gamma$ point
with the additional low-dispersion midgap band passing through it.
These $\sqrt{3}$ and $3$ structures,
which have unit cells that contain $3$ and $9$ unit cells of graphene,
respectively, are somewhat exceptional
because the hydrogen adsorption fails to
open an expected gap in their band structure.
The difference between these and the ordinary cases can be deduced from
the Brillouin zone structures drawn in figure 2.
As shown in figures 2(a) and (b),
related to $\sqrt{3}\times\sqrt{3}R30^\circ$ and $3\times3$, respectively,
both K points of graphene coincide with the $\Gamma$ point of the
hydrogenated graphene for these structures.
As a result there is a four-fold degeneracy at the $\Gamma$ point
in the centre of the Brillouin zone.
On the other hand, for the other structures, $4\times4$ and $5\times5$,
shown in figures 2(c) and (d), the K points of graphene coincide with the K
points of hydrogenated graphene at the corner of the Brillouin zone.
As bandgap opening due to periodic perturbations occur on the boundaries
of the Brillouin zone,
the above observation accounts for the
dependence of bandgap opening on periodicity.
Such a phenomenon can also be seen in alkali metal adsorption,
which have different bonding properties and most notably
adsorb on the hollow sites of graphene instead of the top sites
\cite{farj2009}.
For alkali metals, however, the situation is reversed in that
band-gap opening is absent for the regular cases,
and occurs instead for the periodicities that involve a multiple of 3.

We can now examine band structures for the coverages
discussed above calculated numerically using spin-polarized DFT.
All four band structures, shown in figure 3,
have spin-polarized bands indicating a magnetic moment of
1 $\mu_\mathrm{B}$ per unit cell, with the exchange splittings that decrease
as H concentration is reduced.
Figures 3(a) and (b) for $\sqrt{3}\times\sqrt{3}R30^\circ$ and $3\times3$
coverages show small, but finite, bandgap openings at the $\Gamma$ point.
The shift of the neutrality point and gap opening to the $\Gamma$ point
is in agreement with the analysis
in terms of the Brillouin zones. But
in contrast with the simple TB model which gives zero bandgap, here there is
a small bandgap due to additional symmetry breakings induced by H adsorption.
Figures 3(c) and (d) show the regular cases of $4\times4$ and $5\times5$
coverages where graphene bands open a gap at the K point
and there is a pair of spin-polarized midgap bands. 
The projected densities of states of corresponding to
the $4\times4$ band structure is also shown in figure 4(a).
The energy bands around the gap in figures 3(a) and (b) have an intricate
structure.
The majority up-spin midgap state is degenerate with the top of the
graphene $\pi$ band at $\Gamma$ point below the Fermi energy,
and the same is mirrored for the
minority down-spin midgap state and the bottom of
$\pi^\ast$ band above the Fermi energy.
The origin of this structure must be an additional
symmetry breaking not accounted for in the TB model of (1).

An AB sublattice symmetry breaking can be expected on account of the
formation of spin and charge textures due to hydrogen chemisorption
[see figure 4(b) and reference 16].
We can uncover it by examining the \textit{ab initio} Hamiltonian
matrix \cite{reic2002},
of which we list the most important elements in table~1 \cite{nb1}. 
If we use a tight-binding model based on this table, we can reproduce
the bandgap level structures of figures 3(a) and (b).
The AB symmetry breaking is clearly seen in the spin-dependent on-site energies,
which are tabulated for the 1s orbital of H,
and a set of 2p$_z$ orbitals of C atoms,
indexed with 0, 1, 2, and 3,
which includes the adsorption site and its three nearest neighbours
[they occupy the triangular shape in figure 4(b)].
Note that the even indices refer to the sublattice of the adsorption site,
and the odd indices to the complementary sublattice. 
Setting aside the singular C$_0$ for the time being,
the AB sublattice symmetry breaking
is manifested in the on-site energies for the three C$_1$,
the six C$_2$, and the three C$_3$ sites.
In particular, for up-spin parameters
the energy of C$_2$ is higher than those of C$_1$ and C$_3$.
For the down-spin parameters the sense of symmetry breaking is reversed
with energy of C$_2$ lower than those of C$_1$ and C$_3$.
It is precisely this ordering that lifts the degeneracies of
$\sqrt{3}\times\sqrt{3}R30^\circ$ and $3\times3$ coverages in the manner
shown in figure 3.

Returning to the case of C$_0$,
we note that its on-site energy has been shifted down by more that 2 eV
relative to the other carbon atoms.
We also see that the C$_0$-C$_1$ hopping is suppressed
by nearly 1 eV relative to its
value for clean graphene, which is $-2.96$ eV for the basis set used in our
calculation.
The usual description of the effect of
hydrogen chemisorption on the electronic structure of graphene
is that it removes a p$_z$ orbital from
the $\pi$ system of graphene
\cite{yazy2010, yazy2007}.
As a consequence a quasi-localized state is formed which populates
the complementary lattice.
This in turn gives rise to a midgap energy level near the Fermi energy which
makes the system unstable with respect to spin polarization.
The shifted values of the on-site energy
and the hoppings in table~1 that make C$_0$
different from other C atoms provide quantitative parameters for this
description.

Finally, from table~1
we note that the on-site energy of H is a few electron-volts
below those of the carbon atoms,
which is what we expect when we think in terms of atomic orbitals.
This is another important difference with the semiempirical model of (1),
where $\varepsilon_\mathrm{H}=-0.2$~eV.
The TB model (1), however, provides a better description of a dispersionless
midgap band near the Fermi energy.
The shortcoming of the present
\textit{ab initio} model is that next-neighbor
hoppings and overlap matrices have been neglected.
We may note that
up to third nearest-neighbor hoppings and overlaps had to be included
for an accurate TB description of graphene
\cite{reic2002}.

\begin{figure}
\centering
\includegraphics[width=0.3\linewidth]{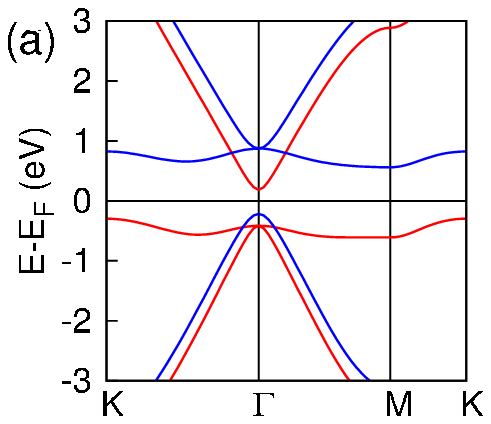}
\includegraphics[width=0.3\linewidth]{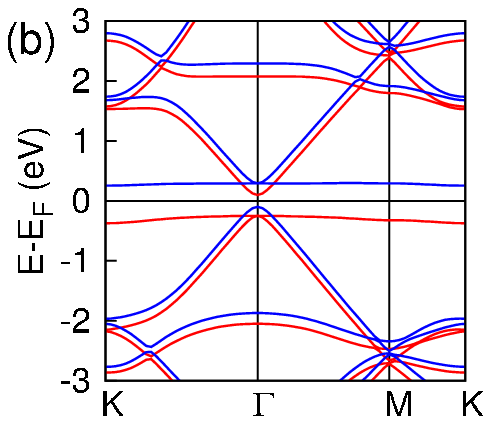} \\
\includegraphics[width=0.3\linewidth]{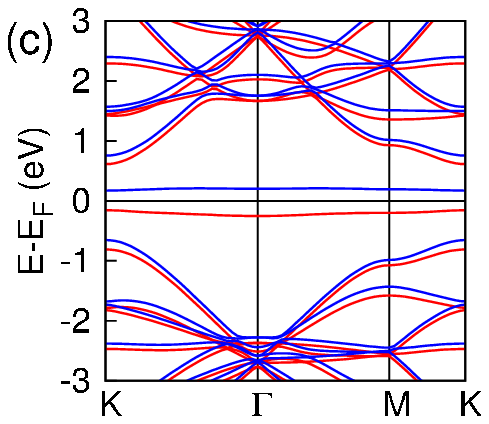}
\includegraphics[width=0.3\linewidth]{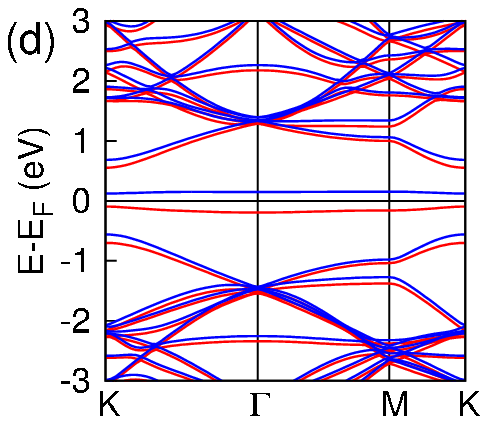}
\caption{\label{fig3} Band structures of hydrogenated graphene shown for a
window surrounding the
neutrality point for (a) $\sqrt{3}\times\sqrt{3}R30^\circ$,
(b) $3\times3$, (c) $4\times4$, and (d) $5\times5$ structures.
The bands are split in pairs due to spin polarization.
In (a) and (b) graphene $\pi$ bands open at $\Gamma$, while in (c) and (d)
they open at K. The points
$\Gamma$, K and M represent the centre, corner and middle of the side of
the hexagonal Brillouin zones of the corresponding structures, respectively.}
\end{figure}

\Table{Hamiltonian matrix elements involving H 1s and C 2p$_z$ orbitals.}
\br
&&\centre{2}{Onsite energies}&&&\centre{2}{Hoppings}\\
\ns
&Equivalent&\crule{2}&&&\crule{2}\\
Atom&sites&$\varepsilon_\uparrow$ (eV)&
$\varepsilon_\downarrow$ (eV)&&
Bond&$\gamma_\uparrow$ (eV)&$\gamma_\downarrow$ (eV)\\
\mr
H&1&$-7.94$&$-7.56$&&H-C$_0$&$-7.02$&$-6.92$\\
C$_0$&1&$-4.80$&$-4.83$&&C$_0$-C$_1$&$-2.17$&$-2.14$\\
C$_1$&3&$-2.48$&$-1.68$&&C$_1$-C$_2$&$-3.02$&$-2.99$\\
C$_2$&6&$-1.76$&$-2.03$&&C$_2$-C$_3$&$-2.91$&$-2.88$\\
C$_3$&3&$-2.21$&$-1.42$&&&&\\
\br
\end{tabular}
\end{indented}
\end{table}

\begin{figure}
\psfrag{(a)}{\bf (a)}
\psfrag{(b)}{\bf (b)}
\centering
\includegraphics[width=0.4\linewidth]{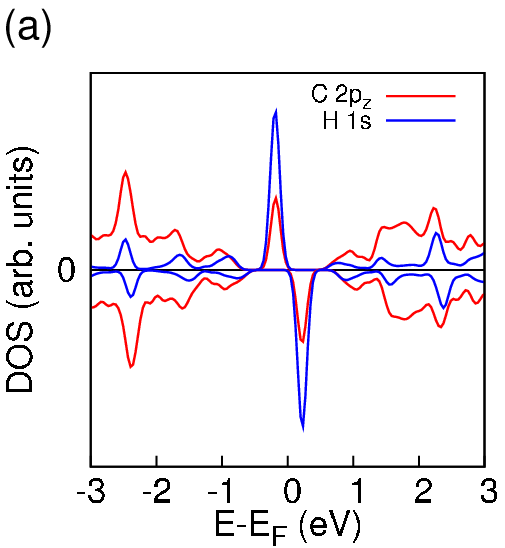} \\ \vspace{2mm}
\includegraphics[width=0.35\linewidth]{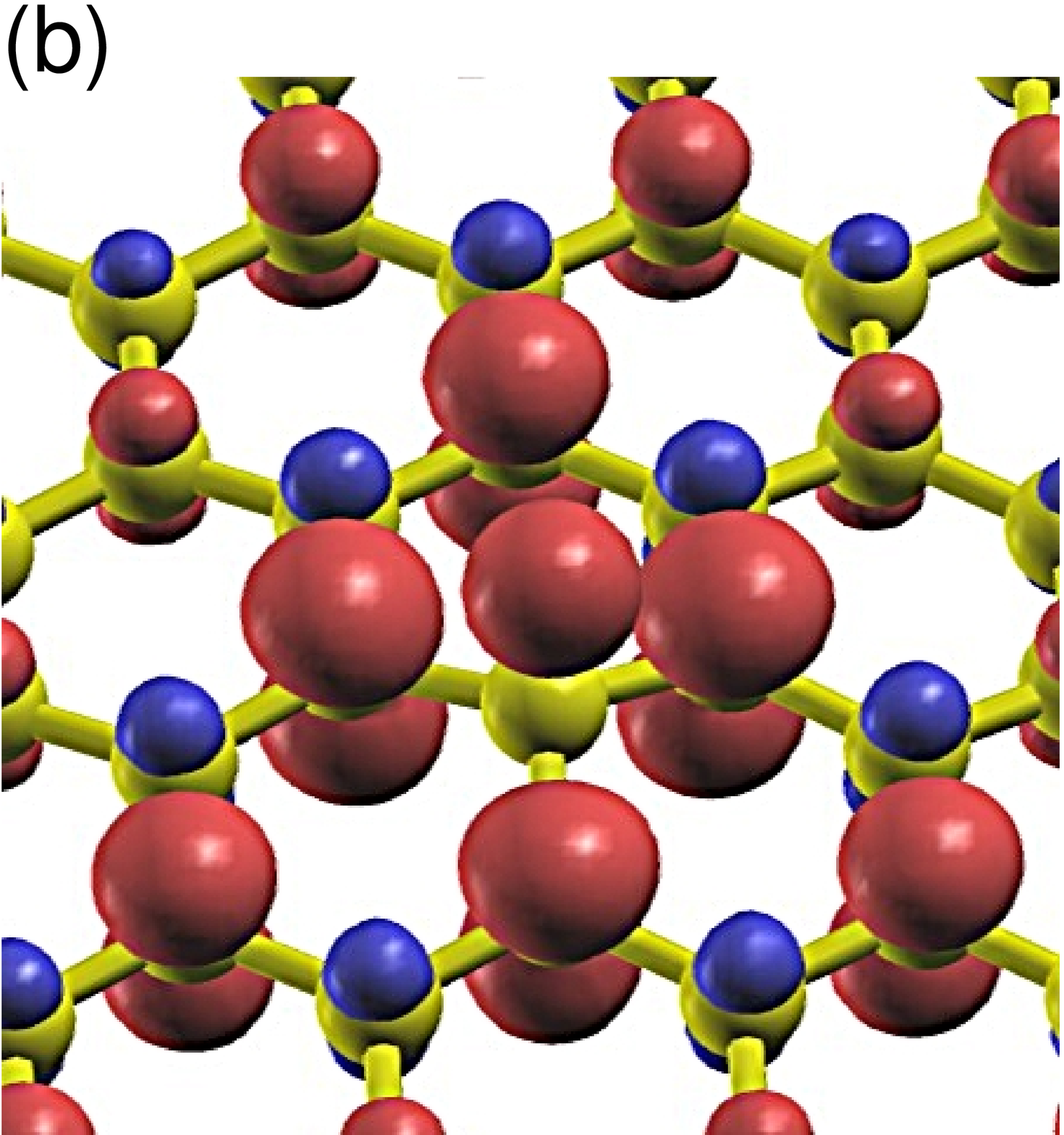}
\caption{\label{fig4}
(a) Spin-polarized projected density of states per atom for $4\times4$
hydrogenated graphene, with up-spin and down-spin parts shown in the upper
and lower halves of the plot, respectively.
(b) Spin density texture caused by H adsorption, showing the typical
$\sqrt{3}\times\sqrt{3}R30^\circ$ texture,
and AB sublattice symmetry breaking.}
\end{figure}

For periodic arrangement of defects it has been shown that the width of
the bandgap is inversely proportional to defect concentration
\cite{mart2010}.
Our calculations provide the data to study this dependence.
Thus we plot the bandgap,
and exchange splittings, defined as the difference
between the spin-polarized midgap bands
measured between the peaks of the density of states \cite{yazy2007},
as a function of number of C atoms in the supercell,
$2n^2$ carbon atoms for $n\times{n}$ coverage,
and find a similar trend of $1/n^2$ for both quantities in figure 5.
Our values of exchange splittings are in favorable agreement with
the value of 0.23 eV for a hydrogen concentration of 0.5\% given in
other calculations
\cite{yazy2007}.

\begin{figure}
\centering
\includegraphics[width=0.5\linewidth]{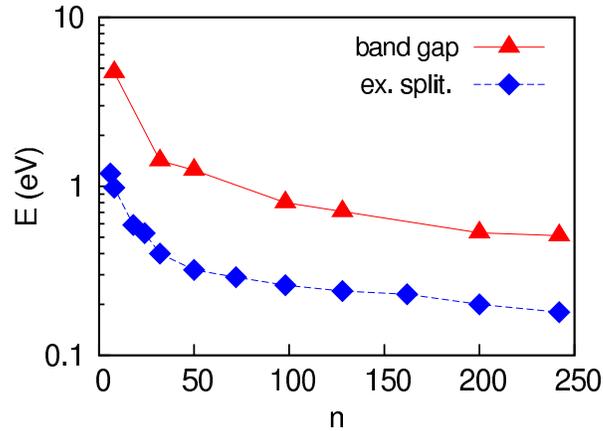}
\caption{\label{fig5}
Band gap openings at K point for a number of structures up to $11\times11$
coverage, 242 atoms,
as a function of the number of carbon atoms in the unit cell, $n$.
Special structures, $3\times3$, etc.,
which have very small bandgaps and do not follow the trend have been left out.
Exchange splittings are plotted for the whole sequence,
and are seen to be about a third of the corresponding bandgaps.
The trend for both curves is $\propto1/n$.}
\end{figure}

\section{Conclusions}
We have made density-functional theory calculations of graphene with
periodic arrangements of hydrogen adsorbates.
Our focus was to explore and explain
the changes made to the electronic band structure of graphene
as a result of hydrogen adsorption.
The important features include bandgap opening and the emergence
of spin-polarized midgap bands,
which we have collectively described as bandgap effects.
Our primary purpose was to examine the dependence of bandgap effects
on hydrogen periodicity, or coverage, within spin-polarized density-functional
theory.
We showed that for periodicities divisible by $3$ or $\sqrt{3}$
bandgap opening has a different behaviour than for $4$, $5$,
and, in general, for periodicities not divisible by 3.
We illustrated that the existence of different classes
has purely geometrical reasons.
In addition, the changes to the band structures have special features,
which we explained on the basis of the AB sublattice symmetry breaking
that is caused by hydrogen adsorption.
A secondary purpose was to gain a deeper understanding of bandgap effects
by examining the \textit{ab initio} Hamiltonian matrix elements.
The tight-binding parameters obtained by this approach reveal the AB
sublattice symmetry breaking, as well as the origins of the midgap bands.
These parameters are physically more meaningful
than those of the simple semiempirical model
and may be developed into a more accurate tight-binding description
of hydrogen adsorption on graphene.

\ack{We would like to thank Professor~H~Rafii-Tabar for helpful comments.}

\section*{References}

\begin{thebibliography}{10}

\bibitem{novo2004}
Novoselov K~S, Geim A~K, Morozov S~V, Jiang D, Zhang Y, Dubonos S~V,
  Grigorieva I~V, and Firsov A~A
2004 {\em Science} {\bf 306} 666

\bibitem{novo2005}
Novoselov K~S, Geim A~K, Morozov S~V, Jiang D, Katsnelson M~I,
  Grigorieva I~V, Dubonos S~V, and Firsov A~A
2005 {\em Nature} {\bf 438} 197

\bibitem{zhan2005}
Zhang Y, Tan Y~W, Stormer H~L, and Kim P
2005 {\em Nature} {\bf 438} 201

\bibitem{wall1947}
Wallace P~R
1947 {\em Phys. Rev.} {\bf 71} 622

\bibitem{cast2009}
Castro~Neto A~H, Guinea F, Peres N~M~R, Novoselov K~S, and Geim A~K
2009 {\em Rev. Mod. Phys.} {\bf 81} 109

\bibitem{schw2010}
Schwierz F
2010 {\em Nature Nanotech.} {\bf 5} 487

\bibitem{elia2009}
Elias D~C~{\it et al}
2009 {\em Science} {\bf 323} 610

\bibitem{balo2010}
Balog R~{\it et al}
2010 {\em Nature Mater.} {\bf 9} 315

\bibitem{habe2010}
Haberer D~{\it et al}
2010 {\em Nano Lett.} {\bf 10} 3360

\bibitem{habe2011a}
Haberer D~{\it et al}
2011 {\em Phys. Rev. B} {\bf 83} 165433

\bibitem{yazy2010}
Yazyev O~V
2010 {\em Rep. Prog. Phys.} {\bf 73} 056501

\bibitem{ni2010}
Ni Z~H~{\it et al}
2010 {\em Nano Lett.} {\bf 10} 3868

\bibitem{kato2010}
Katoch J, Chen J-H, Tsuchikawa R, Smith C~W, Mucciolo E~R, and
  Ishigami M
2010 {\em Phys. Rev. B} {\bf 82} 081417

\bibitem{jelo1999}
Jeloaica L and Sidis V
1999 {\em Chem. Phys. Lett.} {\bf 300} 157

\bibitem{dupl2004}
Duplock E~J, Scheffler M, and Lindan P~J~D
2004 {\em Phys. Rev. Lett.} {\bf 92} 225502

\bibitem{yazy2007}
Yazyev O~V and Helm L
2007 {\em Phys. Rev. B} {\bf 75} 125408

\bibitem{bouk2008}
Boukhvalov D~W, Katsnelson M~I, and Lichtenstein A~I
2008 {\em Phys. Rev. B} {\bf 77} 035427

\bibitem{caso2009}
Casolo S, L\o{}vvik O~M, Martinazzo R, and Tantardini G.~F.
2009 {\em J. Chem. Phys.} {\bf 130} 054704

\bibitem{verg2010}
Verg\'es J~A and de~Andres P~L
2010 {\em Phys. Rev. B} {\bf 81} 075423

\bibitem{wehl2010}
Wehling T~O, Yuan S, Lichtenstein A~I, Geim A~K, and Katsnelson M~I
2010 {\em Phys. Rev. Lett.} {\bf 105} 056802

\bibitem{farj2011}
Farjam M, Haberer D, and Gr\"uneis A
2011 {\em Phys. Rev. B} {\bf 83} 193411

\bibitem{yuan2010}
Yuan S, De~Raedt H, and Katsnelson M~I
2010 {\em Phys. Rev. B} {\bf 82} 115448

\bibitem{garc2010}
Garc\'ia-Lastra J~M
2010 {\em Phys. Rev. B} {\bf 82} 235418

\bibitem{reic2002}
Reich S, Maultzsch J, Thomsen C, and Ordej\'on P
2002 {\em Phys. Rev. B} {\bf 66} 035412

\bibitem{sole2002}
Soler J~M, Artacho E, Gale J~D, Garc\'ia A, Junquera J, Ordej\'on P,
  and S\'anchez-Portal D~S
2002 {\em J. Phys.: Condens. Matter.} {\bf 14} 2745

\bibitem{trou1991}
Troullier N and Martins J~L
1991 {\em Phys. Rev. B} {\bf 43} 1993

\bibitem{perd1996}
Perdew J~P, Burke K, and Ernzerhof M
1996 {\em Phys. Rev. Lett.} {\bf 77} 3865

\bibitem{monk1976}
Monkhorst H~J and Pack J~D
1976 {\em Phys. Rev. B} {\bf 13} 5188

\bibitem{farj2009}
Farjam M and Rafii-Tabar H
2009 {\em Phys. Rev. B} {\bf 79} 045417

\bibitem{nb1}
Obtaining the \textit{ab initio} tight-binding parameters
does not need any extra calculations, because
the Hamiltonian, overlap matrices, etc., are printed by
\textsc{Siesta} to special files,
SystemLabel.HSX in binary format,
and DMHS.nc in NetCDF format.
Our task is only to open either of these
files with appropriate tools and read the selected elements.

\bibitem{mart2010}
Martinazzo R, Casolo S, and Tantardini G~F
2010 {\em Phys. Rev. B} {\bf 81} 245420

\end{thebibliography}

\end{document}